\documentclass[runningheads]{llncs}
\usepackage{graphicx}
\usepackage{multirow}
\usepackage{tikz}
\usepackage{comment}
\usepackage{amsmath,amssymb} 
\usepackage{color,xcolor,colortbl}
\usepackage[pagebackref=true,breaklinks=true,letterpaper=true,colorlinks,bookmarks=false]{hyperref}
\usepackage{subfig}
\usepackage{wrapfig,lipsum,booktabs}
\usepackage{caption}
\usepackage[accsupp]{axessibility}  

\usepackage{algorithm,algorithmic,bm,color}

\usepackage{bm}

\definecolor{lightgray}{gray}{0.95}
\definecolor{color3}{gray}{0.95}
\definecolor{rouse}{rgb}{0.981,0.961,0.941}

\usepackage{subfiles}
\usepackage[normalem]{ulem}
\usepackage{microtype}
\usepackage{url}
\usepackage{booktabs}
\usepackage{amsfonts,xcolor,bbm,mathrsfs,xcolor} 
\usepackage{algorithm,algorithmic,bm,color}
\usepackage{multirow}
\usepackage{wrapfig}
\usepackage{adjustbox}
\usepackage[noadjust]{cite}
\usepackage{graphics, color,  dsfont}
\usepackage{bbding}
\usepackage{makecell}
\usepackage[utf8]{inputenc}
\usepackage{babel}
\usepackage[font=small,labelfont=bf]{caption}

\newcommand{\argmin}{\arg\!\min}

\newcommand{\Cmat}{{\bf C}}
\newcommand{\Dmat}{{\bf D}}
\newcommand{\Emat}[0]{{{\bf E}}}

\newcommand{\Hmat}[0]{{{\bf H}}}
\newcommand{\Imat}{{\bf I}}

\newcommand{\Xmat}{{\bf X}}
\newcommand{\Ymat}[0]{{{\bf Y}}}
\newcommand{\Zmat}{{\bf Z}}

\newcommand{\iv}[0]{{\boldsymbol{i}}}

\newcommand{\mv}[0]{{\boldsymbol{m}}}

\newcommand{\uv}[0]{{\boldsymbol{u}}}
\newcommand{\vv}{\boldsymbol{v}}

\newcommand{\xv}{\boldsymbol{x}}
\newcommand{\yv}{\boldsymbol{y}}

\newcommand{\zv}{\boldsymbol{z}}

\newcommand{\lambdav}[0]{{\boldsymbol{\lambda}} }

\newcommand{\ie}[0]{\emph{i.e.}}
\newcommand{\etc}[0]{\emph{etc.}}

\begin{document}
\pagestyle{headings}
\mainmatter

\title{Ensemble Learning Priors Driven Deep Unfolding for Scalable Video Snapshot Compressive Imaging} 


\titlerunning{Ensemble Learning for Video SCI}
%
\author{Chengshuai Yang \orcidID{0000-0003-2840-5344} 
\and
Shiyu Zhang \orcidID{0000-0001-7111-3895} 
\and
Xin Yuan\Envelope\orcidID{0000-0002-8311-7524}}
\authorrunning{Yang C., Zhang S. and Yuan X.}
%
\institute{School of Engineering, Westlake University, Hangzhou, Zhejiang 310030, China \\
\email{integrityyang@gmail.com ~ \{zhangshiyu, xyuan\}@westlake.edu.cn}
}

\maketitle

\begin{abstract}
Snapshot compressive imaging (SCI) can record a 3D datacube by a 2D measurement and algorithmically reconstruct the  desired 3D information from that 2D measurement. The reconstruction algorithm thus plays a vital role in SCI. Recently, deep learning (DL) has demonstrated outstanding performance in reconstruction, leading to better results than conventional optimization-based methods. 
Therefore, it is desirable to improve DL reconstruction performance for SCI. 
Existing DL algorithms are limited by two bottlenecks: 1) a high-accuracy network is usually large and requires a long running time; 2) DL algorithms are limited by scalability, \ie, a well-trained network cannot generally be applied to new systems. To this end, this paper proposes to use \textbf{ensemble learning priors} in DL to achieve high reconstruction speed and accuracy in a single network. Furthermore, we develop the scalable learning approach during training to empower DL to handle data of different sizes without additional training. Extensive results on both simulation and real datasets demonstrate the superiority of our proposed algorithm. The code and model 
can be accessed at \url{https://github.com/integritynoble/ELP-Unfolding/tree/master}.

\let\thefootnote\relax\footnotetext{\Envelope~Corresponding author.} 

\end{abstract}

\section{Introduction \label{Sec:intro}}

Recently, video snapshot compressive imaging (SCI) \cite{Llull:13,Yuan14CVPR,Chen22Optica} has attracted much attention because it can improve imaging speed by capturing three-dimensional (3D) information from 2D measurement. When video SCI works, multiple frames are first modulated by different masks (in the optical domain), and these modulated frames are mapped into a single measurement. After this, the reconstruction algorithm recovers these multiple frames from single measurement~\cite{Yuan2021_SPM}. At present, the mask can easily be adjusted with a higher speed than the capture rate of the camera~\cite{Hitomi11ICCV,Qiao2020_APLP,2011CVPR_Reddy}. Thus, SCI enjoys the advantages of high speed, low memory, low bandwidth, low power and potentially low cost~\cite{Yuan_2020_CVPR,Yuan2021_TPAMI}.

\begin{wrapfigure}{l}{0.46\textwidth}
  \begin{center}
    \includegraphics[width=0.46\textwidth]{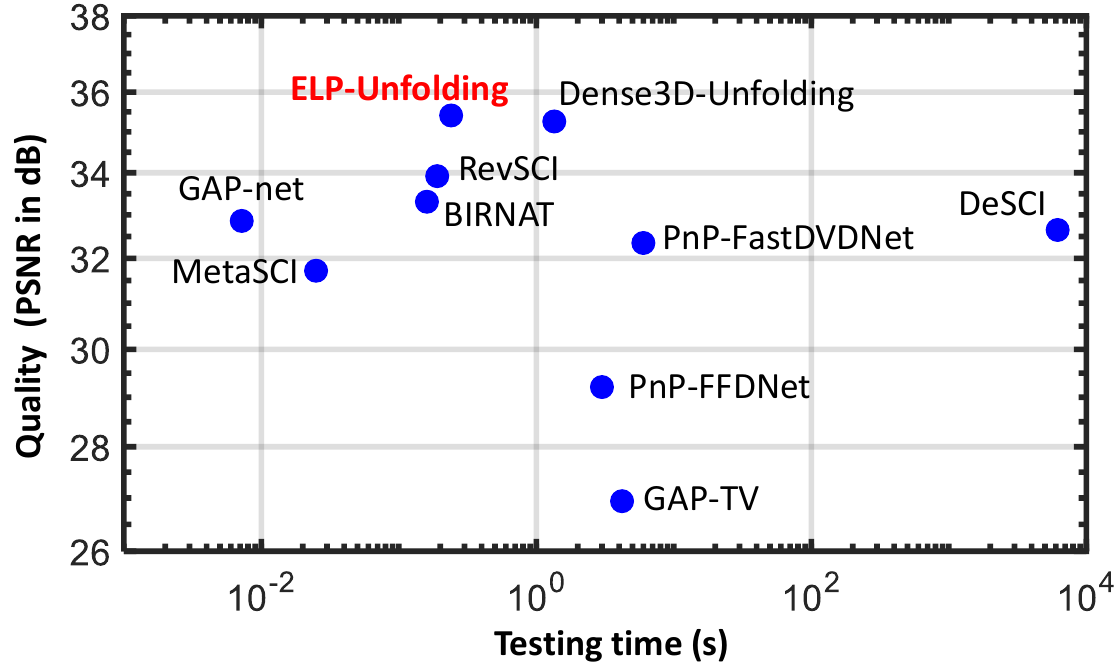}
  \end{center}
	\caption{ Trade-off between quality and testing-time of various algorithms for SCI reconstruction. Our proposed  Ensemble Learning Priors (ELP) unfolding achieves the state-of-the-art results in a short testing time. Besides, after scalable learning, our ELP-Unfolding can be used in different masks and different compression ratios and thus can be applied to various scenes by a single trained model.}
	\label{fig:compare}
\end{wrapfigure}

How to recover the original multiple frames from the single measurement always plays a vital role in SCI. Recently, deep learning reconstruction methods have outperformed traditional iterative reconstruction methods not only in reconstruction accuracy but also in test time~\cite{kulkarni2016reconnet,mousavi2015deep,shi2017deep,yang2016deep,yang2018admm,cheng2020birnat,cheng2021memory,Cheng22_TPAMI}.
But most deep learning methods lack interpretability. 
To increase interpretability, deep unfolding method has been developed, which simulates the iterative algorithm~\cite{gregor2010learning,yang2018admm,zhang2018ista,meng2020gap}. Deep unfolding method adopts iterative framework but replaces traditional denoiser (such  as total variation~\cite{bioucas2007new,li2013efficient} and nonlocal self-similarity~\cite{dong2014compressive,liu2018rank}) with the trained neural network denoiser. So far, the deep unfolding method has achieved the best result for SCI. Among deep unfolding algorithms, GAP-net~\cite{meng2020gap} can use the shortest time (0.0072 s) to achieve 32 dB for PSNR for benchmark dataset. Dense3D-Unfolding~\cite{wu2021dense} achieved the best result (35 dB), though it costs a long time (1.35 s) due to the use of complex 3D convolutional neural networks (CNNs). Thus, {\em the speed and accuracy have not coexisted in one algorithm yet}. What's worse, most of these deep learning algorithms are limited by scalability. To apply the trained model to new systems, the model usually should be trained again. 
Although MetaSCI~\cite{wang2021metasci} can be quickly  applied to new SCI modulations (in spatial but not in temporal dimension), it still requires adaptation (retraining). 

Bearing the above concerns in mind, in order to achieve a higher reconstruction accuracy with a high computing speed, we develop the Ensemble Learning Priors (ELP) unfolding based on 2D-CNN for SCI. Specifically, 2D-CNN can retain fast processing and ensemble learning can increase reconstruction accuracy. Ensemble learning is powerful in achieving reconstruction accuracy and has also achieved state-of-the-art (SOTA) results in a number of  models on other tasks~\cite{zheng2019new,zhou2021domain,pintelas2020special}, due to the fact that mutiple models/priors have complementary advantages over a single model/prior. Fortunately, the deep unfolding algorithm can include many  neural network priors, even if these priors stay at different (iteration) stages. In this paper, we first propose to gather multiple neural network priors in one stage to realize {\bf ensemble learning} for SCI without increasing training time. To further increase the reconstruction accuracy, dense connection is employed in our network, which can help the latter (stage) models learn some useful information from the previous (stage) models. In this manner, our ELP-unfolding can achieve SOTA result, outperform Dense3D-Unfolding~\cite{wu2021dense}, and use a shorter running time. 

Furthermore, to realize the scalability, we develop a scalable learning procedure for SCI. Our method {\em not only has scalability in the spatial dimensions but also in the temporal dimension}. Considering the spatial scalability, we set our ELP-unfolding to be fully convolutional without the multilayer perception (MLP) structure. For temporal dimension scalability, the input of neural network priors is set to have the same channels even for different temporal dimensional scenes. Based on this, our scalable learning method can have the same capability as traditional iteration algorithms, to be applied to different systems.
Specific contributions of our paper are listed as follows:
\begin{itemize}
\item We develop the {\bf ensemble learning prior unfolding} for SCI. ELP unfolding is a general method for inverse algorithms, which can also be applied to other fields, such as single pixel camera~\cite{higham2018deep,radwell2019deep,gibson2017real}, MRI~\cite{akkus2017deep,liu2018applications}, lensless imaging~\cite{antipa2019video,yuan2018parallel,yuan2016slope}, spectral compressive imaging~\cite{Meng2020d,Zheng21_PR_PnP_CASSI,He2021_TIP,cai2021mask,lin2022coarse,hu2022hdnet}, and tomography imaging~\cite{yoo2019deep,dong2019deep,Qiao2021_TCI}.   
\item We first propose the {\bf scalable learning for SCI}. 
After training once, our model can be used in new systems with different modulations or different compression ratios. Besides, scalable learning can achieve better results than PnP algorithm with a fast inference speed.
\item We adopt skip connection techniques in unfolding. In our ELP-unfolding, the skip connection only uses the simple adding and concatenating. By contrast, the Dense3D-Unfolding~\cite{wu2021dense} adopts complex methods such as DFMA (dense feature map adaption) to realize connection. 
\item Our method {\bf achieves SOTA results} for SCI in benchmark dataset based on 2D-CNN, outperforming the 3D-CNN method at a faster inference speed~\cite{wu2021dense} as shown in Fig.~\ref{fig:compare}. 
\end{itemize}

In a nutshell, our ensemble learning priors unfolding has two periods. In the first period, i.e. a single prior period, each stage contains one neural network prior. Afterwards, in the second (ensemble priors) period, each stage contains all previous stage priors in the ensemble manner.

\section{Related Work \label{Sec:Related_work}}

SCI is related to compressive sensing (CS)~\cite{jalali2019snapshot,ZHENG2021_Patterns}, where reconstruction is significantly important as it provides the desired signals (such as images) from the compressed measurements. For CS~\cite{jalali2019snapshot,kulkarni2016reconnet,mousavi2015deep,shi2017deep,meng2020gap}, there are two kinds of reconstruction methods: traditional iterative method and deep learning method. The traditional iteration method contains a lot of iterations and each iteration contains the projection operation and denoising operation (and optionally some other steps). The denoising operation generally determines the performance of one algorithm. For example, total variation~\cite{bioucas2007new,li2013efficient} denoiser has a fast speed but usually can only provide blurry images while the nonlocal self-similarity based denoiser~\cite{dong2014compressive,liu2018rank} can achieve a clearer image but take a long time. Recently, deep learning has shown strong power in reconstructing images~\cite{kulkarni2016reconnet,mousavi2015deep,shi2017deep,yang2016deep,yang2018admm}. At first, deep learning was regarded as a black box and the trained model can get the better images than traditional iterative method at a fast speed. As a black box to train, the trained model will contain measurement matrix (masks) information. Thus, the training model usually can not be applied to new masks (such as a new hardware system). To address this problem, the deep unfolding method for CS has be developed. Deep unfolding method simulates traditional iterative method using a few iterations (stages), each of which has  projection operation and denoising operation. Different from traditional methods, deep unfolding uses a trained neural network as a denoising prior. Therefore, deep learning mainly contributes to denoising in deep unfolding method with little dependence on mask information. The mask information is mainly processed by the projection operation. Thus, deep unfolding algorithms has a strong robustness to a variety of masks~\cite{zhang2018ista,meng2020gap}. Besides, Dense3D-Unfolding~\cite{wu2021dense} obtained SOTA for SCI by combining deep unfolding method and 3D-CNN, but at the cost of slow computation. Though the unfolding method can solve the scalability problem of various masks, the scalability problem of various sizes (both spatial size and temporal size, a.k.a., the compression ratio) still remains in unfolding method. Deep unfolding method still does not have the same scalability as the traditional iterative method.

To address these challenges, in this paper, we develop the ensemble learning priors unfolding for scalable SCI. We use ensemble learning and 2D-CNN to realize high reconstruction accuracy and speed, and develop scalable learning to realize scalability.

\section{Preliminary: Video SCI System \label{Sec:SCI_model}}
\begin{figure}[ht]
	\begin{center}
			\includegraphics[width=0.9\textwidth]{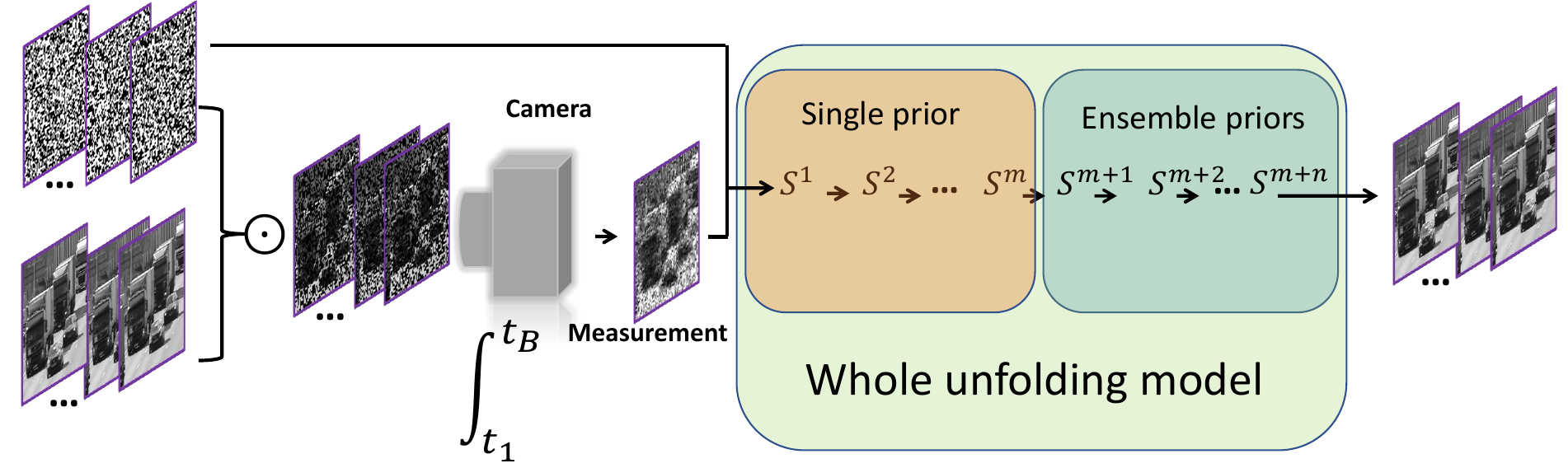}
	\end{center}
	\caption{Principle of Video SCI (left) and our ELP-unfolding (right). \textbf{Left}: the high speed dynamic scene at timestamps $t_1$ to $t_B$, encoded by high-speed variant masks (dynamic coded apertures) and then integrated to a single coded measurement (a compressed image)  $\Ymat$. \textbf{Right}: our whole ELP-unfolding reconstructs the original dynamic scene from the masks $\{\Cmat_1, \dots, \Cmat_B\}$ and the compressed image $\Ymat$, which includes the single prior period in Fig.~\ref{fig:ensemble priors}(a) and ensemble priors period in Fig.~\ref{fig:ensemble priors}(b). $S^m$ represents the $m^{th}$ stage. }
	\label{fig:principle}
\end{figure}
As depicted in Fig.~\ref{fig:principle}, let $\{\Xmat_1, \dots, \Xmat_{B}\}$ denote the discretized video frames at timestamps $\{t_1, \dots, t_B\}$. These video frames are modulated by dynamic coded aperture, a.k.a., the masks $\{\Cmat_1, \dots, \Cmat_B\}$, respectively. The modulated frames are then integrated into a single coded measurement (a compressed image)  $\Ymat$. Here, $\{\Xmat_b\}_{b=1}^B \in {\mathbb R}^{n_x\times n_y\times 
B}$, $\{\Cmat_b\}_{b=1}^B \in {\mathbb R}^{n_x\times n_y\times B}$ and $\Ymat\in {\mathbb R}^{n_x\times n_y}$. This forward model can be written as
 \begin{equation}
\textstyle \Ymat =  \sum_{b=1}^B \Cmat_b\odot \Xmat_b + \Zmat, 
\label{Eq:a}
\end{equation}
 where $\odot$ and $\Zmat\in {\mathbb R}^{n_x\times n_y}$ denote the matrix element-wise product and noise, respectively. Eq.~\eqref{Eq:a} is equivalent to the following linear form
 \begin{equation}
\yv = \Hmat \xv + \zv,  
\label{Eq:b}
\end{equation}
where $\yv={\rm Vec}(\Ymat)\in {\mathbb R}^{n_x n_y}$,$\zv={\rm Vec}(\Zmat)\in {\mathbb R}^{n_x n_y}$ and $\xv={\rm Vec}(\Xmat)=[{\rm Vec}(\Xmat_1)\ts, \dots, \\{\rm Vec}(\Xmat_B)\ts]\ts\in {\mathbb R}^{n_x n_yB}$. Different from traditional compressive sensing~\cite{Donoho06ITT,Candes06ITT,Duarte08SPC}, the sensing matrix $\Hmat$ in~\eqref{Eq:b} has a very special structure and can be written as
 \begin{equation}
\Hmat=[\Dmat_1, \dots, \Dmat_B],
\label{Eq:c}
\end{equation}
where $\{\Dmat_b=diag({\rm Vec}(\Cmat_b))\in {\mathbb R}^{n_xn_y\times n_xn_y} \}_{b=1}^B $are diagonal matrices of masks. Therefore, the compressive sampling rate in SCI is equal
to $1/B$. The reconstruction error of SCI is bounded even when $B>1$~\cite{jalali2019snapshot}.

\section{Our proposed methods~\label{Sec:proposed methods}}

\subsection{Ensemble learning priors unfolding for SCI~\label{Sec:ELP_unfolding_SCI}}

Given the compressed measurement $\Ymat$ and coding pattern $\{\Cmat_b\}_{b=1}^B$ captured by the SCI system, there exist two optimization frameworks to predict the desired high speed frames $\{\Xmat_b\}_{b=1}^B$: penalty function method and augmented Lagrangian (AL) method. The performance of AL method is better than that of the penalty function method, which has been proved in previous work~\cite{li2010efficient,afonso2010augmented,yang2019improving}. Therefore the AL method is adopted here, which is formulated as follows:
\begin{equation}
 \xv= \textstyle\argmin_{\xv}\Phi(\xv)-\lambdav_1^T(\yv - \Hmat \xv)+ 
 \textstyle \frac{\gamma_1}{2} \left\|\yv - \Hmat \xv\right\|_2^2,
\label{Eq:d}
\end{equation}
where $\Phi(\xv),\lambdav_1$ and $ \gamma_1$ denote the prior regularization, Lagrangian multiplier and penalty parameter, respectively.  For convenience, Eq.~\eqref{Eq:d} is further written as
\begin{align}
   \xv= \textstyle \argmin_{\xv}\Phi(\xv)+ \frac{\gamma_1}{2} \left\|\yv - \Hmat \xv-\frac{\lambdav_1}{\gamma_1}\right\|_2^2.
\label{Eq:e}
\end{align}

\begin{figure}
    \centering
    \includegraphics[width=.98\textwidth]{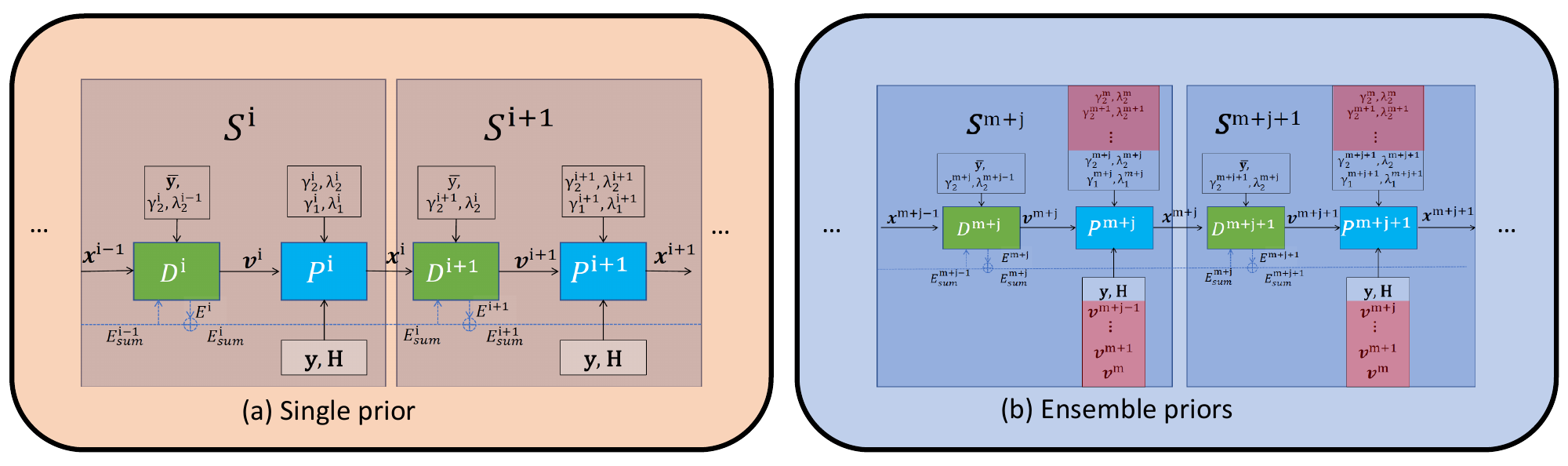}\\
    ~\\
    \caption{(a) Principle of the single prior period. Here, ${D^i}$ represents the $i^{th}$ denoising operation, as in Eq.~\eqref{Eq:j} while ${P^i}$ represents the $i^{th}$ projection operation, as in  Eq.~\eqref{Eq:i}. (b) Principle of ensemble priors period. Here, several denoising results $\vv^{m}...\vv^{m+1}$ are gathered together projection operation.}
    \label{fig:ensemble priors}
\end{figure}

\subsubsection{Single prior.} 
To solve Eq.~\eqref{Eq:e}, an auxiliary variable $\vv$ is introduced. Then Eq.~\eqref{Eq:e} is further written as 
\begin{align}
   \xv= \textstyle\argmin_{\xv}\Phi(\vv)+ 
    \frac{\gamma_1}{2} \left\|\yv - \Hmat \xv-\frac{\lambdav_1}{\gamma_1}\right\|_2^2 ~\text{subject to} ~~\vv =  \xv.
\label{Eq:f}
\end{align}
By adopting alternating direction method of multipliers (ADMM) method~\cite{Boyd11ADMM_y,yang2021high}, Eq.~\eqref{Eq:f} is further written as
\begin{align}
   \xv,\vv= \textstyle \argmin_{\xv,\vv}\Phi(\vv)+ \frac{\gamma_2}{2} \left\|\xv - \vv-\frac{\lambdav_2}{\gamma_2}\right\|_2^2  + \frac{\gamma_1}{2} \left\|\yv - \Hmat \xv-\frac{\lambdav_1}{\gamma_1}\right\|_2^2.
\label{Eq:g}
\end{align}
According to ADMM, Eq.~\eqref{Eq:g} can be divided into the two subproblems and solved iteratively, as shown in Fig.~\ref{fig:ensemble priors}(a)
\begin{align}
    \vv^{i}&=\textstyle\argmin_{\vv}\Phi(\vv) +\frac{\gamma_2^{i}}{2} \left\|\xv^{i-1} - \vv-\frac{\lambdav_2^{i}}{\gamma_2^{i}}\right\|_2^2\label{Eq:h1},\\
   \xv^{i}&=\textstyle\argmin_{\xv} \frac{\gamma_2^{i}}{2} \left\|\xv - \vv^{i}-\frac{\lambdav_2^{i}}{\gamma_2^{i}}\right\|_2^2 + \textstyle\frac{\gamma_1^{i}}{2} \left\|\yv - \Hmat \xv-\frac{\lambdav_1^{i}}{\gamma_1^{i}}\right\|_2^2,
\label{Eq:h2}
\end{align}
where the superscript \emph{i} denotes the iteration index. 

For subproblem $\xv_i$, there exists a closed-form solution, which is called projection operation
\begin{align}
   \xv^{i}=\textstyle(\gamma_2^{i}\Imat +\gamma_1^{i}\Hmat^{T}\Hmat)^{-1}\left[\lambdav_2^{i}+\gamma_2^{i}\vv^{i}+\Hmat^{T}\gamma_1^{i}(\yv -\frac{\lambdav_1^{i}}{\gamma_1^{i}})\right].
\label{Eq:i}
\end{align}
Due to the special structure of {\Hmat}, this can be solved in one shot~\cite{liu2018rank}.

For subproblem $\vv_i$,  Eq.~\eqref{Eq:h1} can be rewritten as
\begin{align}
   \vv^{i}=\textstyle\argmin_{\vv}\Phi(\vv) +\frac{\gamma_2^{i}}{2} \left\|\uv^{i-1} - \vv\right\|_2^2\ ,
\label{Eq:j}
\end{align}
where $\uv^{i-1}=\xv^{i-1}-\frac{\lambdav_2^{i-1}}{\gamma_2^{i}}$. Eq.~\eqref{Eq:j} is a classical denoising problem, which can be solved by denoising prior such as TV, wavelet transformation, denoising network, \etc  In this paper, denoising network prior is adopted as shown in Fig.~\ref{fig:unet}.



\subsubsection{Ensemble priors.}
In every stage of unfolding, the denoising prior has different parameters and thus plays different roles in removing noise, even these priors have the same structure. To take full use of different denoisers among different stages, these priors after $\mv$ stages are gathered together to perform projection operation to produce $\xv$. Therefore, Eq.~\eqref{Eq:h1} and Eq.~\eqref{Eq:h2} in ensemble priors period, as shown in Fig.~\ref{fig:ensemble priors}(b), becomes
\begin{align}
    \vv^{m+j}=\textstyle\argmin_{\vv}\Phi(\vv_1) +\frac{\gamma_2^{m+j}}{2} \left\|\xv^{m+j-1} - \vv-\frac{\lambdav_2^{m+j-1}}{\gamma_2^{m+j}}\right\|_2^2,
\label{Eq:ensemble1}
\end{align}
and
\begin{align}
    \xv^{m+j}=&\textstyle\argmin_{\xv} \frac{\gamma_2^{m}}{2} \left\|\xv -   \vv^{m}-\frac{\lambdav_2^{m}}{\gamma_2^{m}}\right\|_2^2 +\textstyle\frac{\gamma_2^{m+1}}{2} \left\|\xv - \vv^{m+1}-\frac{\lambdav_2^{m+1}}{\gamma_2^{m+1}}\right\|_2^2 + \cdots
    \nonumber \\
   &  \quad + \textstyle\frac{\gamma_1^{m+j}}{2} \left\|\yv - \Hmat \xv-\frac{\lambdav_1^{m+j}}{\gamma_1^{m+j}}\right\|_2^2.
\label{Eq:ensemble2}
\end{align}
For subproblem $\vv_i$, Eq.~\eqref{Eq:ensemble1} can still adopt the same denoising prior form as in single-prior period. For subproblem $\xv_i$, there is a slightly difference because of ensemble
\begin{align}
   &\xv^{m+j}=[(\gamma_2^{m}+\gamma_2^{m+1}+\cdots+\gamma_2^{m+j})\Imat+\Hmat^{T}\gamma_1^{m+j}\Hmat]^{-1}\nonumber \\
   &\quad\left[\lambdav_2^{m}+\gamma_2^{m}\vv^{m}+\cdots +\textstyle\lambdav_2^{m+j}+\gamma_2^{m+j}\vv^{m+j}+\Hmat^{T}\gamma_1^{m+j}(\yv -\frac{\lambdav_1^{m+j}}{\gamma_1^{m+j}})\right].
\label{Eq:ensemblex}
\end{align}
Last but not least, the Lagrangian multipliers $\lambdav_1^{i}$ and $\lambdav_2^{i}$ are updated by 
\begin{align}
   \lambdav_1^{i}&=\lambdav_1^{i-1}-\gamma_1^{i}(\yv - \Hmat \xv^{i-1}), \label{Eq:lambdav_1} \\
  \lambdav_2^{i}&=\lambdav_2^{i-1}-\gamma_2^{i}(\xv - \vv^{i-1}). \label{Eq:lambdav_2} 
\end{align}
Besides, the $\gamma_1^{i}$ and $\gamma_2^{i}$ are trained with the denoising prior parameters at every stage.

In our method, the whole algorithm body should be divided into two parts: a single prior period and an ensemble priors period, because the first several stages can only provide rough estimates. If the priors in the first several stages are coupled to the latter stages, the poor performance of the first several priors will worsen the whole algorithm performance. There are 13 stages in our algorithm, the first 8 stages are single prior periods and latter 5 stages are ensemble priors periods. It is noted that there are 6 priors in last stage. As we can see in Eq.~\eqref{Eq:ensemble2} and Eq.~\eqref{Eq:ensemblex}, there exist six $\vv$'s if $j=5$.


\subsubsection{Denoising prior structure.} 
As shown in Fig.~\ref{fig:unet}, U-net~\cite{Unet_RFB15a_y} is used as the backbone for denoising prior, {which we adopt from} FastDVDnet~\cite{tassano2020fastdvdnet}, but here we remove batch normalization and quadruple the depth; this  means that the channels for three different features are 128, 256 and 512, respectively. Thus, the training parameters of our proposed ELP-unfolding mainly consists of these 13 U-net structures. {More details can be found in the supplementary materials (SM).} Following~\cite{tassano2020fastdvdnet,zhang2018ffdnet},  the penalty parameter $\gamma_2^{i}$ is expanded to a noise map as part of the input. To help denoising, the normalized measurement  $\overline{\Ymat}$ is also added to the input~\cite{wang2021metasci,cheng2021memory,wu2021dense}, which is defined as 
\begin{align}
   \overline{\Ymat}=\textstyle\Ymat\oslash\sum_{b=1}^B \Cmat_b,
\label{Eq:k}
\end{align}
where $\oslash$ represents the matrix element-wise division. Therefore, the input consists of noise map $\gamma_2^{i}$, normalized measurement $\overline{\Ymat}$ and $\xv^{i-1}-\frac{\lambdav_2^{i-1}}{\gamma_2^{i}}$, and the output is $\vv^{i}$. Besides, dense connection is employed in the denoising prior network design.

\noindent\begin{minipage}{0.48\textwidth}
    \includegraphics[width=1\textwidth]{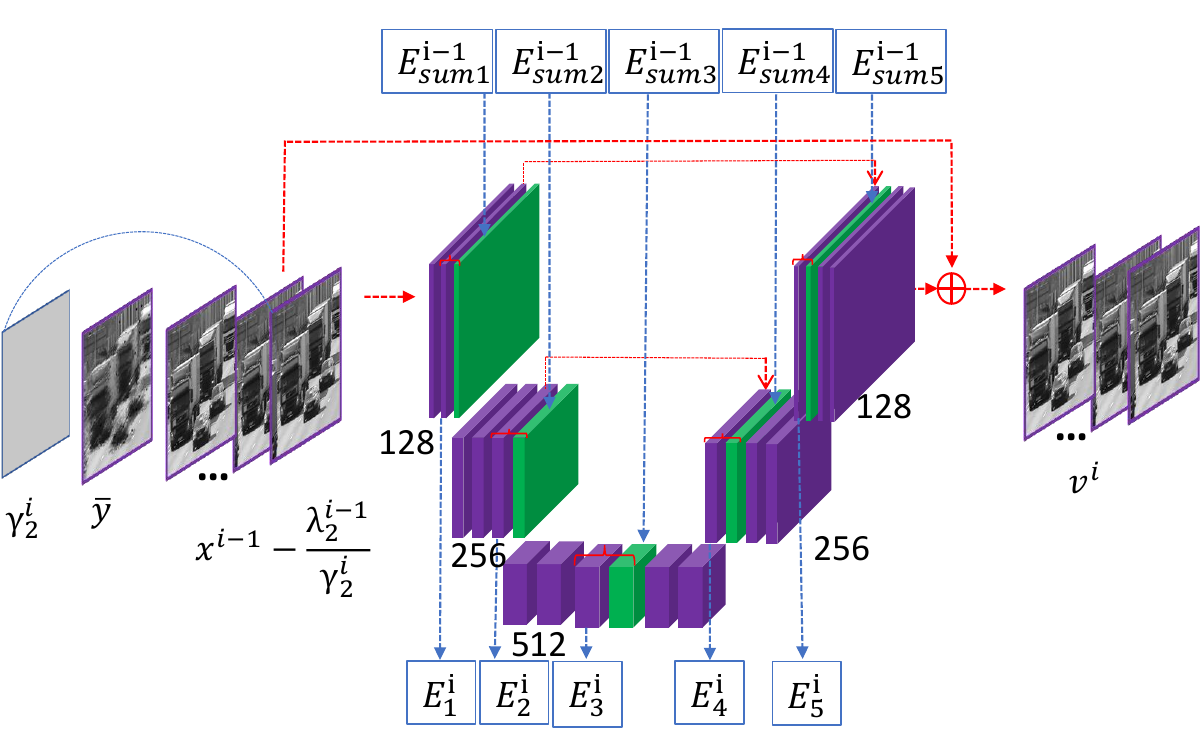}\\
    
	\captionof{figure}{\small Denoising prior structure based on U-net~\cite{Unet_RFB15a_y}. To realize connection, sum feature $\Emat^{i-1}_{sumj}$ from previous priors is coupled into current prior and current feature $\Emat^{i}_j$ is used to help form next sum feature $\Emat^{i}_{sumj}$.}
	\label{fig:unet}
\end{minipage}
~~
\begin{minipage}{0.48\textwidth}
\begin{algorithm}[H]\scriptsize 
	\caption{\small ELP-unfolding for SCI Reconstruction}
	\begin{algorithmic}[1]
		\REQUIRE $\Hmat$, $\yv$, $\overline{\Ymat}$, $\{\gamma_1^{0}$, $\dots$, $\gamma_1^{m+n}\}$, $\{\gamma_2^{0}$, $\dots$, $\gamma_2^{m+n}\}$.
		\STATE Initial $\vv^{0}$ = $\mathbf{0}$, $\lambdav_1^{0}$ = $\mathbf{0}$, $\lambdav_2^{0}$ = $\mathbf{0}$.
		\STATE Update $\xv^{0}$ by Eq.~\eqref{Eq:i} 
		\STATE  \emph{\% single prior period} 
		\FOR{$\mathbf{i}$ = 1, ... , m}
		\STATE Update $\vv^{i}$ by Eq.~\eqref{Eq:j}, $\lambdav_2^{i}$ by Eq.~\eqref{Eq:lambdav_2}. 
		\STATE Update $\lambdav_1^{i}$ by Eq.~\eqref{Eq:lambdav_1}, $\xv^{i}$ by Eq.~\eqref{Eq:i}.
		\ENDFOR
		\STATE  \emph{\% ensemble priors period} 
		\FOR{$\mathbf{k}$ = m+1, ... , m+n}
		\STATE Update $\vv^{k}$ by Eq.~\eqref{Eq:j}, $\lambdav_2^{k}$ by Eq.~\eqref{Eq:lambdav_2}.  
		\STATE Update $\lambdav_1^{k}$ by Eq.~\eqref{Eq:lambdav_1}, $\xv^{k}$ by Eq.~\eqref{Eq:ensemblex}. 
		\ENDFOR
	\end{algorithmic}
	\label{algo:ELP_unfolding}
\end{algorithm}
\end{minipage}
\subsubsection{Dense connection for unfolding.}
In traditional unfolding method, the connection between two stages are $\vv$ and $\uv$, that is $\xv^{i-1}-\frac{\lambdav_2^{i-1}}{\gamma_2^{i}}$, which have a small number of temporal dimensions. Therefore, most latent information in U-net structure cannot be transferred between {different priors}. To break this bottleneck, the skip connection technique is used here. As shown in Fig.~\ref{fig:unet}, in the $\iv^{th}$ {prior}, the feature $\Emat^{i}_j$ and feature  $\Emat^{i-1}_{sumj}$ operate in the latent space of U-net structure as a whole feature. Besides, the feature $\Emat^{i}_j$ will add to $\Emat^{i-1}_{sumj}$ to form $\Emat^{i}_{sumj}$, that is, $\Emat^{i}_{sumj}=\Emat^{i-1}_{sumj}+\Emat^{i}_j$.

By re-ordering the updating equations, we summarize the entire algorithm in Algorithm~\ref{algo:ELP_unfolding}.

\subsection{Scalable learning for SCI~\label{Sec:scalable learning}}
Existing deep learning methods usually have limited scalability, \ie, one trained model can only be applied to one system with specific masks and compression ratio $B$. When the scene data size changes, the new corresponding model usually needs to be trained again. The most recent MetaSCI~\cite{wang2021metasci} can quickly be applied to a new model but also demands new adaptation process. {In addition, MetaSCI adaptation is limited in space but not suitable for time (compression ratio).} Even some deep learning methods that are independent of multi-layer perception, such as Dense3D-Unfolding, can be applied to different spatial size cases, but they have no temporal scalability. They must be trained again for new applications with different temporal dimensions $B$.

To address this problem, we develop scalable learning for SCI. This scalable learning has scalability not only in the spatial dimension but also in the temporal dimension. Specifically, to ensure spatial scalability, we only employ the convolutional neural network, ignoring MLP; to ensure temporal scalability, we train a scalable frames model within a certain number of frames, which is  the maximum  frames ${M}$. During training, the number of frames (smaller than $M$) is randomly chosen; ${M}$ is also the number of channel in denoising networks. In most cases, the original data should be repeatedly rearranged several times to satisfy the frames number ${M}$. When ${M}$ is not an integer multiple of the frame number of dynamic scene, only the first several frames of the original data are used in the last arranging process. 

Even though the maximum temporal size needs to be pre-set, the new maximum temporal model can conveniently use  the previous different maximum temporal models as the pre-trained model to speed up the training process.

\begin{figure}[t]
  \begin{center}
    \includegraphics[width=0.7\textwidth]{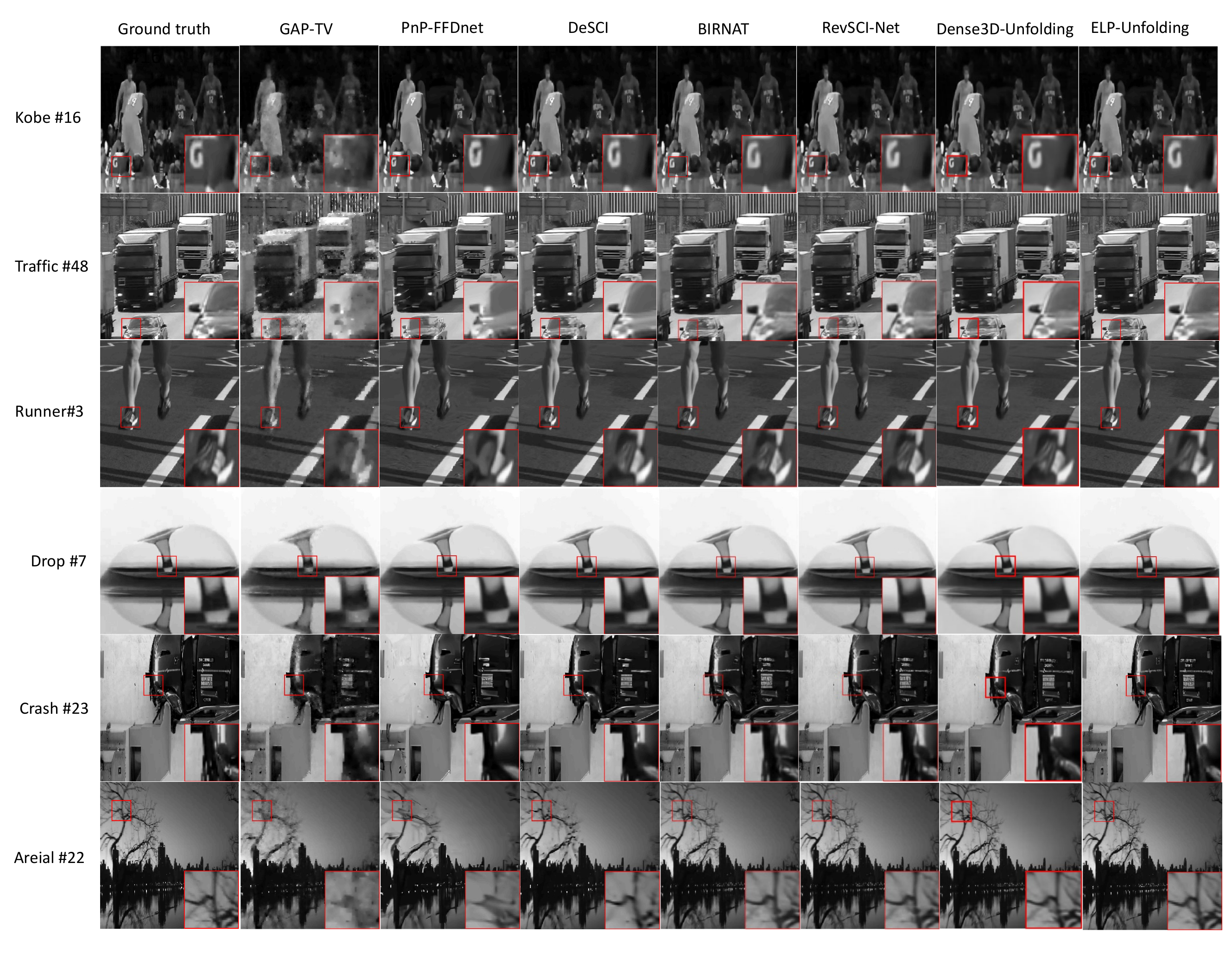}
  \end{center}
	\caption{\small Selected reconstruction results of benchmark dataset by GAP-TV~\cite{Yuan16ICIP}, DeSCI~\cite{liu2018rank}, PnP-FFDNet~\cite{Yuan_2020_CVPR}, RevSCI~\cite{cheng2021memory} and the proposed ELP-unfolding (Please zoom-in to see details).}
	\label{fig:benchmark}
\end{figure}

\subsection{Training~\label{Sec:Training}}

Given the measurement $\Ymat$ and masks $\{\Cmat_b\}_{b=1}^B$, our ELP-unfolding can generate $\{\hat{\Xmat}_b\}_{b=1}^B \in {\mathbb R}^{n_x\times n_y\times B}$. 
The mean square error (MSE) is selected as our
loss function, expressed as
\begin{align}
   \ell_{MSE}=\textstyle \frac{1}{SBn_xn_y}\sum_{s=1}^S\sum_{b=1}^B\left\|\Xmat_b - \hat{\Xmat}_b\right\|_2^2,
\label{Eq:loss}
\end{align}
where $\Xmat_b$ is ground truth and $S$ is batchsize.


We use PyTorch ~\cite{paszke2019pytorch} to train our model on an NVIDIA A40 GPU. For all training processes, we
adopt the Adam optimizer ~\cite{kingma2014adam} with a mini-batch size of 3 and a spatial size of 256 ${\times}$ 256. 
We also adopt a pre-training strategy. The whole training process has two periods. Firstly, 8 stages with a single prior model are trained as pretrained parameters. Secondly, the whole ELP-unfolding with the pretrained parameters, is then trained, with 13 stages, 6 ensemble-priors in the last stage. And the first 8 stages just contains a single prior in each stage. Besides, the former 8 stages in the entire ELP-unfolding match the pretrained model very well, completely adopting the pretrained parameters. 
The latter 5 stages priors adopt the same last stage parameters in the pretrained model.

Regarding the learning rate, we adopt the same strategy for these two training periods. The difference lies in the initial learning rate. For the first (pretrained) period, the initial learning rate is set to  $\text{1}{\times}10^{-4}$. For the second (ELP-unfolding) period, the initial learning rate is set to $\text{2}{\times}10^{-5}$. After the first five epochs, the learning rate decays a factor of 0.9 every 15 epochs. Besides, for the first (pretrained) period, the total number of epoch is 200 and training time is about 8 days. For the second  period, the total number of epoch is 320 and training time is about 13 days.

In this paper we used above training strategies to train three models, namely benchmark model, scalable model and real data model. 

\section{Experiment}
\label{sec:exp}
\subsection{Training Dataset}
We used DAVIS2017~\cite{pont20172017} dataset with a resolution of 480${\times}$894 (480p) as our training dataset for all experiments. Video clips with spatial size of 256 ${\times}$ 256 are randomly cropped from this training dataset.

\begin{table}[t]
	\caption{\small Benchmark datasets: the average results of PSNR in dB (left entry in each cell) and SSIM (right entry in each cell) and run time per measurement in seconds by different algorithms on 6 benchmark datasets.}
	\centering
	\resizebox{.95\textwidth}{!}
	{
	\begin{tabular}{c|cccccc|c|c}
		\hline
		Algorithm& \texttt{Kobe} & \texttt{Traffic} & \texttt{Runner} & \texttt{Drop} & \texttt{Crash} & \texttt{Aerial} & Average &  Run time (s) \\
		\hline
		GAP-TV~\cite{Yuan16ICIP}          & 26.92, 0.838 & 20.66, 0.691 & 29.81, 0.895 & 34.95, 0.966 & 24.48, 0.799 & 24.81, 0.811 & 26.94, 0.833 &  4.2 (CPU) \\
		\hline
		{DeSCI~\cite{liu2018rank}} & 33.25,  0.952 &  28.71, 0.925 &  38.48,  0.969 & 43.10, 0.992 & 27.04, 0.909 & 25.33, 0.860 & 32.65, 0.934 & 6180 (CPU)\\
		\hline
		PnP-FFDNet~\cite{Yuan_2020_CVPR}      & 30.33, 0.925 & 24.01, 0.835 & 32.44, 0.931 & 39.68, 0.986 & 24.67, 0.833 & 24.29, 0.820 & 29.21, 0.888 &  3.0 (GPU) \\
		\hline
		PnP-FastDVDnet~\cite{Yuan2021_TPAMI}   & 32.73, 0.947 & 27.95, 0.932 & 36.29, 0.962 & 41.82, 0.989 & 27.32, 0.925 & 27.98, 0.897 & 32.35, 0.942 & 6 (GPU) \\
		\hline
		BIRNAT~\cite{cheng2020birnat}   &  32.71,  0.950 &  29.33,  0.942 & 38.70, 0.976 & 42.28,  0.992 & 27.84,  0.927 & 28.99, 0.927 & 33.31, 0.951 & 0.16 (GPU) \\
		\hline
		GAP-Unet-S12~\cite{meng2020gap}   &  32.09,  0.944 &  28.19,  0.929 & 38.12, 0.975 & 42.02,  0.992 & 27.83,  0.931 & 28.88, 0.914 & 32.86, 0.947 & {\bf0.0072 (GPU)} \\
		\hline
		Meta-SCI~\cite{wang2021metasci}   &  30.12, 0.907 &  26.95, 0.888 & 37.02, 0.967 & 40.61, 0.985 & 27.33, 0.906 & 28.31, 0.904 & 31.72, 0.926 & 0.025 (GPU) \\
		\hline
		RevSCI~\cite{cheng2021memory}   &  33.72,  0.957 &  30.02,  0.949 & 39.40, 0.977 & 42.93,  0.992 & 28.12,  0.937 & 29.35, 0.924 & 33.92, 0.956 & 0.19 (GPU) \\
		\hline
		Dense3D-Unfolding~\cite{wu2021dense}   & {\bf 35.00}, {\bf 0.969} & {\bf 31.76}, {\bf 0.966} & 40.03, 0.980 & 44.96, {\bf 0.995} & 29.33,  0.956 & 30.46, 0.943 & 35.26, 0.968 & 1.35 (GPU) \\
		\hline
		ELP-Unfolding (Ours)  & 34.41, 0.966 & 31.58, 0.962 & {\bf 41.16}, {\bf 0.986} & {\bf 44.99}, {\bf 0.995} & {\bf 29.65}, {\bf 0.960} & {\bf 30.68}, {\bf 0.943} & {\bf 35.41}, {\bf 0.969} & 0.24 (GPU)  \\
		\hline
	\end{tabular}
    }
	\label{Tab:benchmark}
\end{table}

\subsection{Benchmark datasets for SCI}

Kobe, Traffic, Runner, Drop, Crash, and Aerial are the Benchmark datasets for SCI\cite{Yuan2021_TPAMI}, where the data-size is 256${\times}$256${\times}$8, \ie $n_x$=$n_y$=256, $B$=8. Based on these datasets, we compare our ELP-unfolding with a special temporal size of 8 to other SOTA algorithms, including GAP-TV~\cite{Yuan16ICIP}, DeSCI~\cite{liu2018rank},  PnP-FFDNet~\cite{Yuan_2020_CVPR}, PnP-FastDVDnet~\cite{Yuan2021_TPAMI}, BIRNAT~\cite{cheng2020birnat}, GAP-Unet-S12~\cite{meng2020gap}, Meta-SCI~\cite{wang2021metasci}, RevSCI~\cite{cheng2021memory}, Dense3D-Unfolding~\cite{wu2021dense}. 
The results are summarized in Table~\ref{Tab:benchmark}.
As we can see, iterative algorithms including PnP based algorithms (GAP-TV, DeSCI, PnP-FFDNet, PnP-FastDVDnet) provide inferior results at a slow speed (more than one second). Deep learning algorithms can achieve better result in a short running time (usually less than 1 second).

For direct comparison of deep learning algorithms, Table~\ref{Tab:three DL} shows the results of top three algorithms, namely, RevSCI, Dense3D-Unfolding and ours.  Although Dense3D-Unfolding has achieved the best results before, it costs a long time to test (1.35 s). Our ELP-unfolding algorithm not only  achieves better result than Dense3D-Unfolding, but also saves test time (costing 0.24 s). 
\begin{wraptable}{r}{5cm}
	\caption{\small The comparison of top three algorithms: time, memory for training one batch and reconstruction accuracy (PSNR).}
	\centering
	\vspace{3mm}
	\scalebox{0.7}
	{ 
	\begin{tabular}{|c|c|c|c|}
		\hline
		 & \texttt{Time} & \texttt{Memory} & \texttt{PSNR}\\
		\hline
		RevSCI~  & 0.19 s& Flexible & 33.92 dB\\
		\hline
		Dense3D-Unfolding~  & 1.35 s& 28.7 G & 35.26 dB\\
		\hline
		Our method~  & 0.24 s & 12.5 G & 35.41 dB\\
		\hline
	\end{tabular}
	} 
	\label{Tab:three DL}
\end{wraptable}
For visualization purpose, we also present some images in Fig.~\ref{fig:benchmark}, from the zoom areas we can see that our ELP-unfolding provides much clearer images with sharper edges and more
abundant details than other algorithms, even the Dense3D-Unfolding (Crash). We also believe that by  adopting 3D-CCN, ELP-unfolding can achieve even better results.

\begin{figure}[t]
	\begin{center}
        \includegraphics[width=1\linewidth]{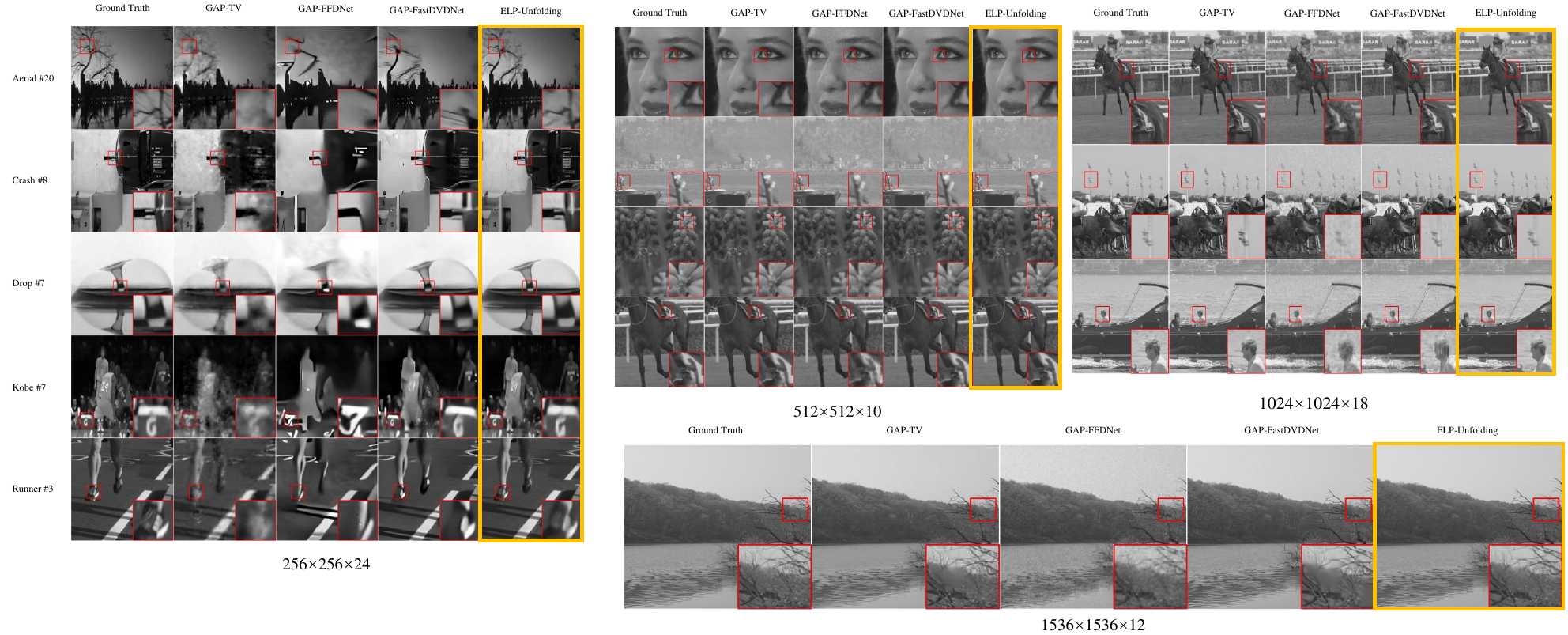}
	\end{center}
	\caption{\small Scalability: Selected results by GAP-TV,  PnP-FFDNet, PnP-FastDVDnet and our ELP-unfolding with various spatial sizes and compression ratios.}
	\label{fig:scalable}
\end{figure}

\subsection{Scalable datasets for SCI}


To verify the scalability of our ELP-unfolding method, we trained one model to test four different size datasets: 256${\times}$256${\times}$24, 512${\times}$512 ${\times}$10, 1024 ${\times}$1024${\times}$18 and 1536${\times}$1536${\times}$12. The latter three datasets are cropped from the Ultra Video Group (UVG) dataset~\cite{mercat2020uvg} in the same way as in Meta-SCI~\cite{wang2021metasci}. The former dataset is also the benchmark. but the compression ratio $B$ is now set to 24. {Because previous deep learning algorithms (including Meta-SCI) cannot scale for different compression ratios, traditional iteration algorithms including} GAP-TV, PnP-FFDNet and PnP-FastDVDnet are chosen as baselines.


\begin{figure}[htbp!]
	\begin{center}
		\includegraphics[width=0.495\linewidth]{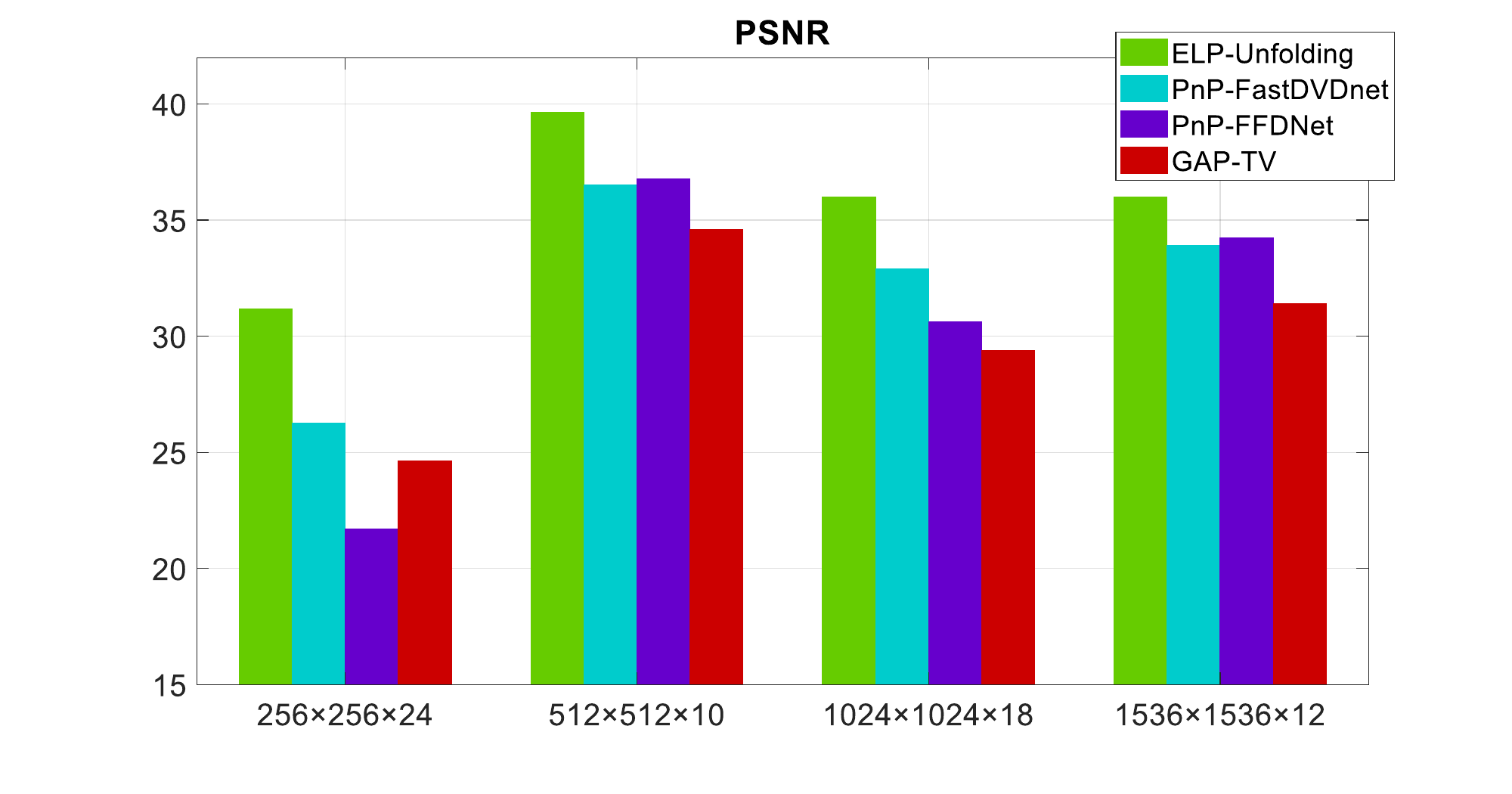}
        \includegraphics[width=.495\linewidth]{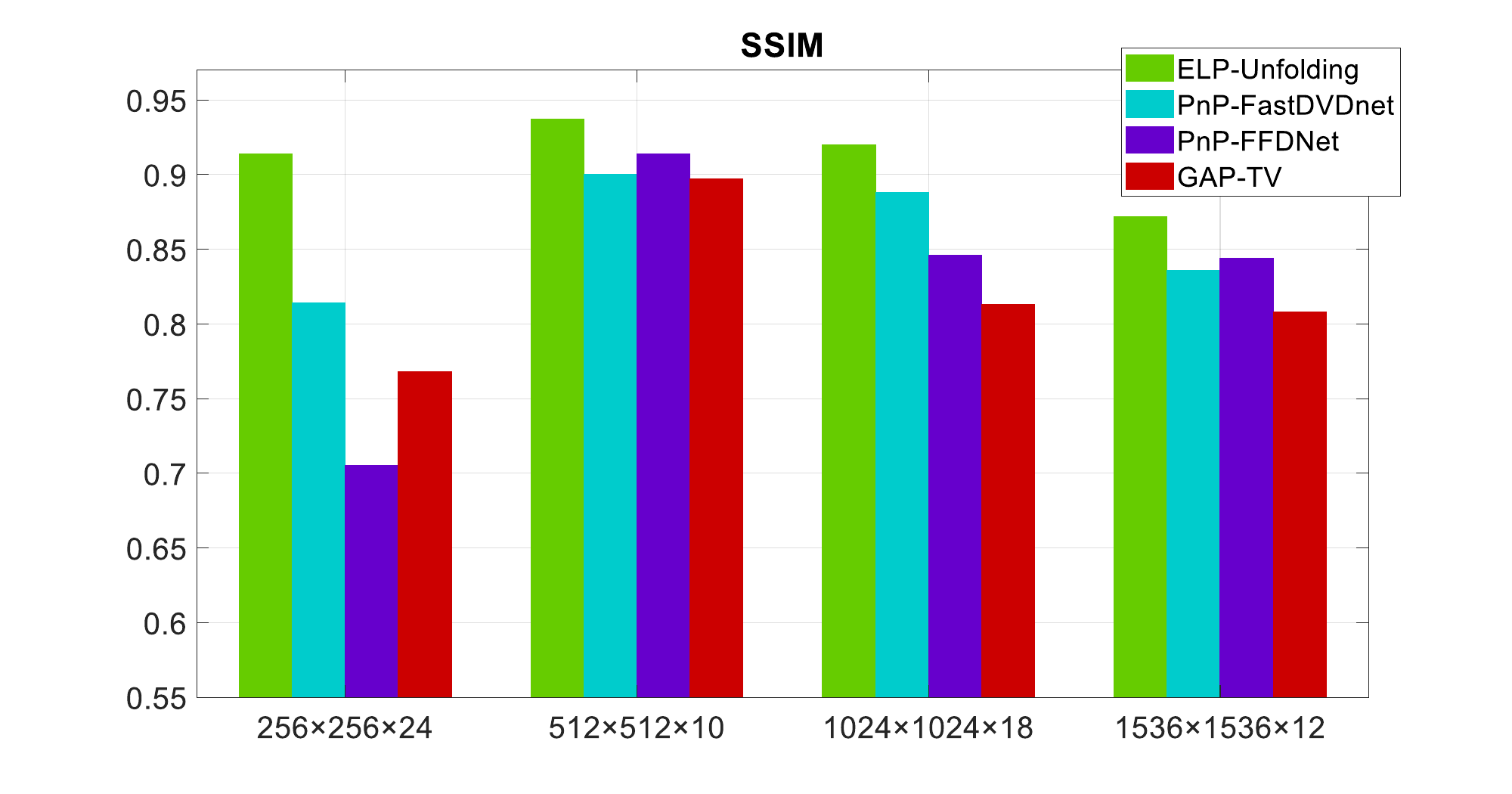}
	\end{center}
	\caption{\small Scalability: Reconstruction results by GAP-TV,  PnP-FFDNet, PnP-FastDVDnet and the proposed ELP-unfolding with various spatial sizes and compression ratios.}
	\label{fig:bar_compare}
\end{figure}

It can be noticed from Fig.~\ref{fig:bar_compare} that  {these algorithms yeild worse results than ELP-unfolding meanwhile cost a longer time} (details in SM). In the case of 1536${\times}$1536${\times}$12, PnP-FFDNet is able to get good results as ELP-unfolding. However, it is unstable and gets the worst results in the case of 256${\times}$256${\times}$24.  Fig.~\ref{fig:scalable} shows some selected images with much sharper boundaries and fewer artifacts reconstructed by ELP-unfolding than other algorithms. Please refer to the reconstructed videos in SM.

\subsection{Ablation Study}
\label{sec:exp_ablation} 
In our ELP-unfolding model, the single prior period contains 8 stages, ensemble
priors period contains 5 stages and thus the whole model contains 13 stages. Stage 9 has two priors to deal with projection operation while stage 10 has three priors and so on and so forth. In the end, in stage 13, there are six priors. 

{Focusing on the number of stages in Table~\ref{Tab:different stages priors}, we can see that the more stages one model has, the better result the model can achieve. But when the number of stages reaches 13, the reconstruction accuracy can not be improved any more. Regarding the priors, by adopting ensemble learning priors strategy, the 6 priors (with 13 stages) model can still improve reconstruction accuracy. Besides, the ensemble learning model always behaves better than its single prior counterpart in the same number of stage case. For instance, in the 9-stage model, two priors in the last stage always leads to better results than the single prior counterpart.}

{Next, we consider a more complicated structure. Specifically, we use 2 priors in stage 2 and 3 priors in stage 3 and so on and so forth. For a fair comparison, we also use a 13stage model. The result of this complicated model is called `Ensemble all' in Table~\ref{Tab:different situation1}. We can observe that even though the model is more complicated, it cannot lead to better results than our proposed structure, because the first several stages  only  provides  rough  estimates and the poor performance of first several priors can deteriorate the whole algorithm performance if coupled to the latter stages. In addition, the `Ensemble no' in Table~\ref{Tab:different situation1} denotes a single prior used in all stages, and the same for the 13 stages in Table~\ref{Tab:different stages priors}. This model can lead to decent results but not as good as ensemble priors structure. After comparison, we set the 6 priors model as our final  ELP-unfolding, the results of which are also shown in 
`Integrating all' in Table~\ref{Tab:different situation1}.}

\begin{table*}[t]
	\caption{\small Ablations. Average PSNR and SSIM for different setups in simulation. }
	\hspace{6mm}\subfloat[\small `m-n' means m stages model has n priors in the last stage. \label{Tab:ensemble priors}]{ 
		\hspace{7mm}\scalebox{0.7}{
			\begin{tabular}{|c c|c c|c c|}
	    \hline
		\texttt{6 single stages}& \texttt{} & \texttt{7 single stages}& \texttt{}& \texttt{8 single stages}& \texttt{}\\
		\hline
		'7-2' 32.69 & '7-1' 30.71 & '8-2' 32.82 & '8-1' 30.98& '9-2' 33.24 & '9-1' 31.25\\
		\hline
		'9-4' 32.92 & '9-1' 31.25 & '9-3' 33.03 & '9-1' 31.25& '11-4' 33.33 & '11-1' 31.45\\
		\hline
		'11-6' 33.06 & '11-1' 31.45 & '11-5' 33.19 & '11-1' 31.45& '13-6' 33.46 & '13-1' 31.75\\
		\hline
		'aver' 32.89 & 'aver' 31.14 & 'aver' 33.13 & 'aver' 31.23& 'aver' 33.34 & 'aver' 31.48\\
		\hline
	\end{tabular}}}\hspace{2mm}
	\subfloat[\small Different stages and ensemble priors in the last stage \label{Tab:different stages priors}]{ 
		\scalebox{0.5}{
			\begin{tabular}{|c|c|c|c|c|c|c|c||c|c|c|c|c|}
		\hline
		\texttt{1 stage}& \texttt{3 stages} & \texttt{5 stages}& \texttt{7 stages} & \texttt{8 stages} & \texttt{9 stages} & \texttt{11 stages} & \texttt{13 stages} & \texttt{1 prior} & \texttt{2 priors} & \texttt{4 priors} & \texttt{6 priors} & \texttt{8 priors}\\
		\hline
		31.21, 0.926   & 33.29, 0.953 & 34.33, 0.964 & 34.50, 0.965 & 34.83, 0.966 & 34.92, 0.967 & 35.11, 0.968 & 35.07, 0.968 &34.83, 0.966   & 34.98, 0.967 & 35.15, 0.968 & 35.41, 0.969 & 35.34, 0.969  \\
		\hline
	\end{tabular}}}\hspace{6mm}
	\subfloat[\small Running 13 stages in different situations.\label{Tab:different situation1}]{
		\hspace{6mm}\scalebox{0.8}{
			\begin{tabular}{|c|c|c|c|c|}
		\hline
		\texttt{Ensemble all}& \texttt{Ensemble no} & \texttt{Part training-set}& \texttt{Removing connection}& \texttt{Integrating all}\\
		\hline
		34.97, 0.967  & 35.09, 0.968 & 34.73, 0.966 & 34.77, 0.966 & 35.41, 0.969 \\
		\hline
	\end{tabular}}}\hspace{2mm}
	\hspace{6mm}\subfloat[\small `7-1' means 7 stages 1 prior while `13-6'means 13 stages 6 priors. \label{Tab:different situation}]{ 
	\hspace{6mm}	\scalebox{0.8}{
		\begin{tabular}{|c|c|c|c|}
		\hline
		\texttt{7-1 w/o connection}& \texttt{7-1 w/ connection} & \texttt{13-6 w/o connection}& \texttt{13-6 w/connection}\\
		\hline
		34.23, 0.961  & 34.85, 0.967 & 34.77, 0.966& 35.41, 0.969 \\
		\hline
	\end{tabular}}}
	\label{tab:ablations}
\end{table*}

\noindent{\bf{Effect of dense connection.}}
To verify the effect of the dense connection in ELP-unfolding, we make the comparisons with and without dense connection in the 7 stages model (`7-1') and 6 priors model (`13-6'). The 7 stages model has seven stages but each stage contains only one prior, while the 6-prior model is the full model in our paper that achieves SOTA results. As shown in Table~\ref{Tab:different situation}, removing the dense connection will lower the performance of unfolding algorithms including the single prior model and the ensemble prior model, because the information transmitted between priors is limited. It should be noticed that our dense connection operation is simple, only consisting of adding and concatenating, instead of complex operations such as the dense feature map adaption in Dense3D-Unfolding~\cite{wu2021dense}. Thus our ELP-unfolding provides a simple strategy (dense connection) to improve the performance of deep unfolding.

\noindent{\bf{Effect of ensemble priors.}}
Table~\ref{Tab:different stages priors} can't completely reflect the effect of ensemble priors, because we adopt a large model by using connection technique and wide channels (512 channels in unet middle layer) to get SOTA accuracy to outperform Dense3D-Unfolding, which leaves little room for ensemble priors improvement. In most circumstances, it is unnecessary to use such big models. Thus, we use the normal 128 channels and remove connection technique to display the effect of ensemble priors, as shown in Table~\ref{Tab:ensemble priors}. As we can see, the ensemble priors method can improve the reconstruction accuracy of PNSR by more than 1.75 dB on average. Besides, ensemble priors method doesn't increase memory  to train and time to test. Thus, ensemble priors have a huge advantage in the field of deep unfolding.

\noindent{\bf{Effect of training dataset.}}
Training dataset plays a key role in performance of deep learning algorithms, and ELP-unfolding is no exception. We verify this by using part of the training dataset, \ie the dataset in DAVIS2017 that only trains on 480p videos, but does not include the test dataset and test challenge dataset. The results are shown in `Part training-set' in Table~\ref{Tab:different situation1}. By comparing `Part training-set' and `Integrating all', we find that reducing the amount of training set hurts the performance of ELP-unfolding. 

\begin{figure}
\centering
\includegraphics[width=0.8\linewidth]{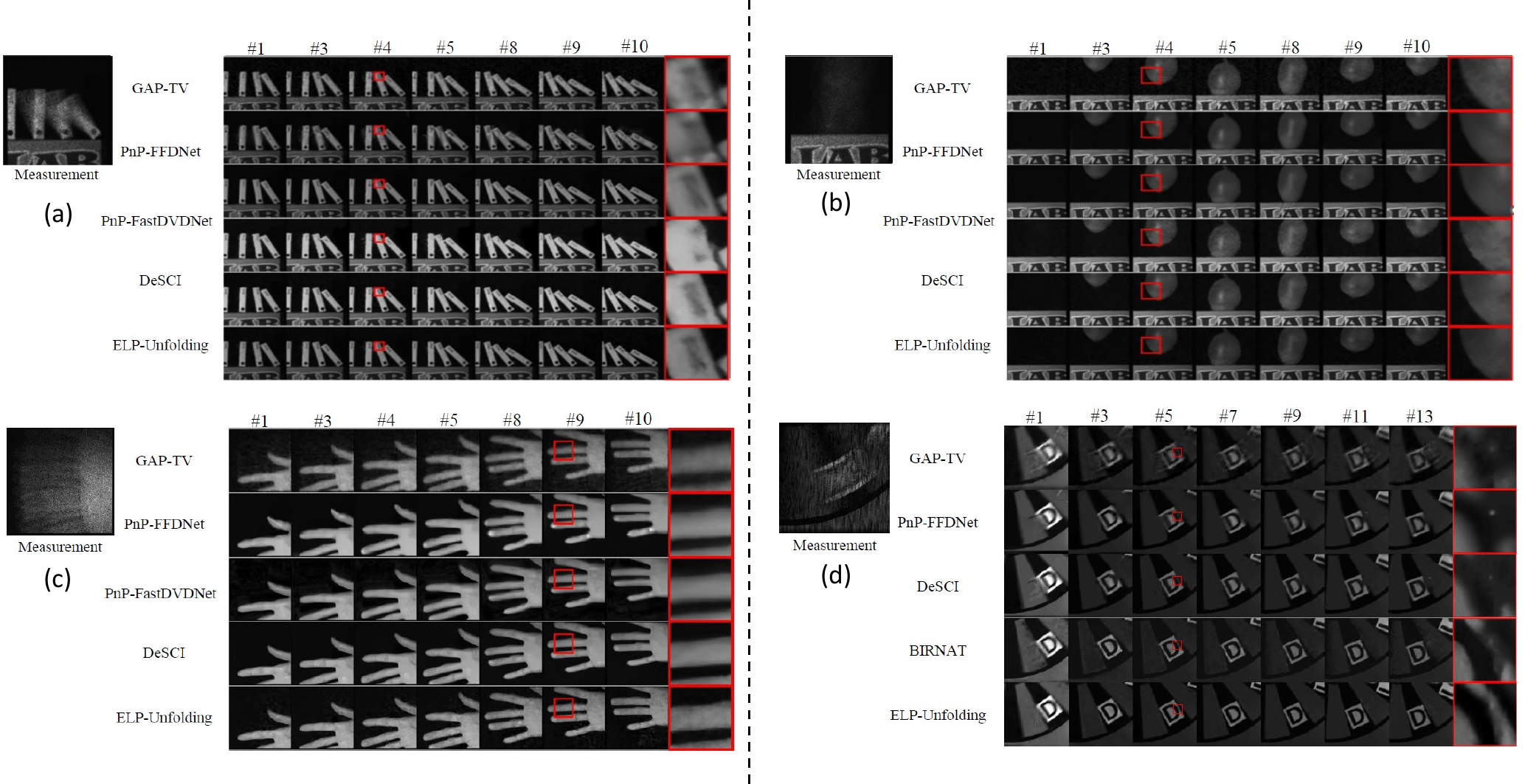}\\
~~\\
\caption{Real data  \texttt{duomino} (a, 512\texttimes512\texttimes10), \texttt{waterballon} (b, 512\texttimes512\texttimes10),   
\texttt{hand} (c, 512\texttimes512\texttimes10) and  \texttt{chop} (d, 256\texttimes256\texttimes14) reconstructed from a compressed measurement.}
\label{fig:domino}
\end{figure}

\section{Real datasets for SCI}
We now apply the proposed ELP-unfolding to real datasets, namely chopwheel~\cite{Llull:13}, waterBalloon~\cite{Qiao2020_APLP}, duomino~\cite{Qiao2020_APLP} and hand~\cite{Yuan2021_SPM}. Because of the unavoidable measurement noise, it is more challenging to reconstruct real measurements. The size of the Chopwheel data is 256${\times}$256${\times}$14, while the size of the other three datasets is 512${\times}$512${\times}$10. From Fig.~\ref{fig:domino}, we can see that our method can generate more apparent contours while reducing artifacts and ghosting. What's more, previous deep learning algorithms didn't succeed in reconstructing hand because of the big noise in this data. Our ELP-unfolding firstly obtains the hand reconstruction by deep learning. Thus, we can only show the comparison with traditional iteration algorithms such as GAP-TV, PnP-FFDNet and DeSCI. Therefore, we can conclude that in practical applications, our method is powerful in reconstructing high-speed scenes.  The relative videos  can be seen in SM.

\section{Conclusions and Future Work}
Inspired by ensemble learning and iterative based optimization algorithm, we develop ensemble learning priors unfolding for scalable snapshot compressive imaging. Our ELP-unfolding algorithm has achieved state-of-the-art results in a short running time. Besides, we have firstly proposed the scalable function for SCI, not only in the spatial dimension but also in the temporal dimension.

To further improve the reconstruction accuracy, we will consider combining 3D-CNN with ELP-unfolding. Besides, to reduce the testing time and the parameters of neural network, a distilling method will be employed. 
We believe that our proposed ELP-unfolding framework can also be used for other inverse problems such as image CS, spectral compressive imaging, and so on~\cite{xue2022block,Zhang22Optica,Yuan_16_OE,Yuan16AO}. 

\subsubsection{Acknowledgements:} 
We would like to thank the Research Center for Industries of the Future (RCIF) at Westlake University, Westlake Foundation (2021B1501-2) and the funding from Lochn Optics.


\clearpage
\bibliographystyle{splncs04}
\bibliography{reference.bib}
\end{document}